\documentclass[a4paper]{eptcs}

\usepackage[utf8]{inputenc}
\usepackage{amsfonts}
\usepackage{graphicx}
\usepackage[usenames,dvipsnames]{xcolor}
\usepackage{multirow}
\usepackage{wrapfig} 

\newcommand{\pt}{\textsc{ProofTool}}
\newcommand{\oldpt}{\texttt{prooftool}}

\newcommand{\LOUI}{\textsc{LOUI}}
\newcommand{\IDV}{\textsc{IDV}}
\newcommand{\CERES}{\textsc{CERES}}
\newcommand{\GAPT}{\textsc{GAPT}}
\newcommand{\IVY}{\textsc{IVY}}
\newcommand{\CLI}{\textsc{CLI}}
\newcommand{\TAP}{\textsc{TAP}}
\newcommand{\ProverN}{\textsc{Prover9}}
\newcommand{\prooftrans}{\textsc{prooftrans}}
\newcommand{\Vampire}{\textsc{Vampire}}
\newcommand{\RegSTAB}{\textsc{RegSTAB}}
\newcommand{\Yes}{\ensuremath{\checkmark}}
\newcommand{\No}{ }

\newcommand{\ceresxml}{\emph{Ceres XML }}
\newcommand{\simplexml}{\emph{Simple XML }}
\newcommand{\hlks}{\emph{Handy LKS }}

\title{\pt: a GUI for the \GAPT{} Framework\thanks{Partially supported by the project I383 of the Austrian Science Fund and the Vienna PhD School of Informatics.}}
\author{Cvetan Dunchev$^1$ \and Alexander Leitsch$^1$ \and Tomer Libal$^1$ \and Martin Riener$^1$ \and Mikheil Rukhaia$^1$ \and Daniel Weller$^2$ \and Bruno Woltzenlogel-Paleo$^1$ \\
\institute{$^1$ Institute of Computer Languages (E185), \\
$^2$ Institute of Discrete Mathematics and Geometry (E104), \\
Vienna University of Technology}
\email{\{cdunchev,leitsch,shaolin,riener,mrukhaia,weller,bruno\}@logic.at}
}
\def\titlerunning{\pt: a GUI for the \GAPT{} Framework}
\def\authorrunning{C.Dunchev, A.Leitsch, T.Libal, M.Riener, M.Rukhaia, D.Weller, B.Woltzenlogel-Paleo}

\begin{document}
\maketitle

\begin{abstract}
This paper introduces {\pt}, the graphical user interface for the General Architecture for Proof Theory (\GAPT) framework. Its features are described with a focus not only on the visualization but also on the analysis and transformation of proofs and related tree-like structures, and its implementation is explained. Finally, {\pt} is compared with three other graphical interfaces for proofs.
\end{abstract}

\section{Introduction}

\GAPT{}\footnote{\GAPT{} homepage: \url{http://www.logic.at/gapt/}} (General Architecture for Proof Theory) is a framework that aims at providing data-structures, algorithms and user interfaces for analyzing and transforming formal proofs.
\GAPT{} was conceived to allow general tools for creating, processing, displaying and transforming structured proofs and one of its first goals was to
replace the \CERES{} system\footnote{\CERES{} homepage: \url{http://www.logic.at/ceres/}}~\cite{ijcar2010,ESARM2008,KEAPPA2008}
and expand its scope beyond the original focus on cut-elimination by resolution for first-order logic~\cite{Baaz2000}. Through a more flexible and succint implementation based on basic data structures for simply-typed lambda calculus and for generic sequent calculus style proofs, in the hybrid functional object-oriented language Scala~\cite{Oderski2010}, this goal has already been achieved: generalizations of the \CERES{} method (cut-elimination by resolution) to proofs in higher-order logic~\cite{Hetzl2011} and to schematic proofs~\cite{Schemata}, as well as methods for structuring and compressing proofs, such as Herbrand sequent extraction~\cite{HLWWP08c,BrunoMSC} and cut-introduction~\cite{HLW12} have been implemented in \GAPT{}.

The \GAPT{} system provides a {\em command line interface} (\CLI) that allows the user to access all functionality of the system, e.g. to create and manipulate proofs and a {\em graphical user interface} called \pt{}\footnote{Both binaries are available in the Download section on the \GAPT{} homepage; the current download links at the time of publication are \url{http://www.logic.at/gapt/gapt-cli-1.4} and \url{http://www.logic.at/gapt/prooftool-1.4}. }. The \CLI{} is very flexible, because it is built on top of Scala's REPL (Read-Evaluate-Print Loop) Interpreter\footnote{Scala's REPL Interpreter: \url{http://www.scala-lang.org/node/2097}}; it is not suited for the visualization of proofs, though, since proofs usually are large, contain specialized mathematical symbols and have a tree-like structure that is inconvenient to read as pure text.

Convenient proof visualization, {\pt}'s main goal, is achieved by means of several features (e.g. tree rendering, scrolling, hiding and highlighting, LaTeX rendering), which are described in Section~\ref{sec:Visualization}.  But {\pt} is more than just a proof viewer. As explained in Section~\ref{sec:ProofMining}, most of \GAPT{}'s features for analyzing and transforming proofs are already available within {\pt}, and we follow the policy of adding new features of \GAPT{} to {\pt} continuously, soon after they reach a stable stage of development.

The fact that {\pt} is more than a proof viewer and needs to support the variety of structural proof theory algorithms allowed in \GAPT{}
is the main reason why simply reusing existing proof viewers, such as \LOUI~\cite{LOUI} and \IDV~\cite{IDV}, was not a suitable option. Another benefit of the close connection between {\pt} and \GAPT{} is that even \GAPT{}'s more exotic objects (e.g. schematic proofs) can be easily displayed and manipulated in {\pt}. It would be hard, if not impossible, to achieve the same functionality on such objects using an external proof viewer. Nevertheless, a comparison between these different systems is available in Section~\ref{sec:conclusion}.
A last but nonetheless important aim is to allow the proof theory community to extend {\GAPT}'s algorithms and data structures. This allows immediate access to the relevant {\pt} features.


\section{Preliminaries}\label{sec:pre}

Most kinds of objects that can be displayed by and manipulated within {\pt} are very specific to the proof mining algorithms implemented in \GAPT{}, and especially to the \CERES{} algorithm of cut-elimination by resolution~\cite{Baaz2000,Baaz2006a}. Among the data structures used by \CERES{} and displayable by {\pt}, the most prominent are:
\begin{itemize}
\item \textbf{Struct}: an unsatisfiable formula obtained from a proof with cuts; when seen as a tree, it has the same branching structure as the proof with cuts.
\item \textbf{Characteristic clause set}: the clause set obtained by transforming the struct into clause form.
\item \textbf{Projection}: a cut-free part of the proof with cuts; each projection corresponds to one clause of the characteristic clause set.
\item \textbf{Resolution Refutation}: a refutation of the characteristic clause set; when combined with the projections, an essentially cut-free proof is obtained.
\item \textbf{Herbrand sequent}: a propositionally valid and quantifier-free sequent made of formulas that are instances of the formulas present in the end-sequent of a cut-free proof; it serves as a compact representation of the first-order content of a proof.
\end{itemize}

\subsection{Proof Schemata}

\emph{Formula schemata} were introduced and investigated in~\cite{Aravantinos2009, Aravantinos2011}. A subclass called {\em regular schemata} was identified and shown decidable, and a tableau calculus STAB was defined and implemented~\cite{regstab-ijcar2010}. A \emph{sequent calculus proof schema} is a primitive recursively specified infinite sequence of sequent calculus proofs, where sequents are multisets of formula schemata. {\pt} is able to display a proof schema as a first-class object, with no need to instantiate it to a particular element of the infinite sequence of proofs it represents. This is made possible with \emph{proof links}: a special axiom rule that is responsible for the recursion. A schematic \CERES{} method for eliminating cuts from proof schemata has been developed~\cite{Schemata} and all its required data structures (e.g. schematic structs, schematic characteristic clause sets, schematic refutations and schematic projections) can be displayed in {\pt} as well.

\section{Supported Input/Output Formats}\label{sec:inputs}

There are several input formats parsed by the \GAPT{} system but the most important ones are the \hlks language, the \ceresxml format and a less restrictive, so called \simplexml format. They will be described below.

The \GAPT{} system supports the \texttt{cnf} and \texttt{fof} subset of the TPTP format~\cite{SS98} and the \IVY{} proof checker format~\cite{ivy}. TPTP exporting is used to communicate with external theorem provers, the \IVY{} format is used to import proofs created by \ProverN.

In practical applications, the input files quickly reach sizes over one MB. Therefore compression was implemented in the form of the {\em .gz} file reader. {\pt} supports parsers for {\em .lks}, {\em .xml} and {\em .ivy} files in gzipped form as well.

In {\pt} there are several exporters for objects of the \GAPT{} system, most importantly {\LaTeX} and {\em .pdf} exports. All objects from {\pt} can be exported directly into {\em .pdf} files and additionally, the proofs and clause sets can be exported into {\em .tex} files as well.

\subsection{Ceres XML}


 \ceresxml is a file format used by the \CERES{} system.
 It can be used to encode DAG-like proofs in a second-order language. Since the GAPT system can deal with proofs which fall
 outside of this language (namely proofs in full higher-order logic), {\ceresxml} is intended to be replaced by a more
 general format in the future. Still, the format has been successfully used in our experiments with first- and second-order proofs,
 and for this reason we include its description here.
 The {\ceresxml} format is mainly described by a Document Type Definition (DTD), which fixes the set of XML files which are
 considered as {\ceresxml}. The DTD
 of the format is quite detailed and restrictive; a part of it is shown in Figure~\ref{fig:ceres_dtd}.

\begin{figure}[h]
\centering
\begin{tabular}{|c|}
\hline \\
\begin{minipage}{5.5in}
\begin{verbatim}
<!ELEMENT proofdatabase (definitionlist,axiomset,proof*,
			  sequentlist*,variabledefinitions)>
<!ELEMENT proof (rule)>
<!ATTLIST proof
          symbol	CDATA	#REQUIRED
          calculus	CDATA	#IMPLIED	>
<!ELEMENT rule	(sequent,(rule|prooflink)*,
		 substitution?,lambdasubstitution?)>
<!ATTLIST rule
          symbol	CDATA	#IMPLIED
          type  	CDATA	#REQUIRED
          param 	CDATA	#IMPLIED	>
<!ELEMENT prooflink	EMPTY>
<!ATTLIST prooflink
          symbol	CDATA	#REQUIRED	>
<!ELEMENT sequentlist	(sequent+)>
<!ATTLIST sequentlist
          symbol	CDATA	#REQUIRED	>
<!ELEMENT sequent	(formulalist,formulalist)>
<!ATTLIST sequent
          projection	CDATA	#IMPLIED	>
<!ENTITY % atomformula	'(constantatomformula|variableatomformula)'>
<!ENTITY % formula
  '(formulavariable|conjunctiveformula|quantifiedformula|
    secondorderquantifiedformula|%atomformula;)'>
\end{verbatim}
\end{minipage} \\ \\
\hline
\end{tabular}
 \caption{A piece of the DTD file of \ceresxml} \label{fig:ceres_dtd}
 \end{figure}

The DTD of the {\ceresxml} format allows a partial correctness check of proofs in XML format:
For example, if an XML file conforms to the DTD, then all the formulas occuring in it are
well-formed. On the other hand, since the actual calculus rules are not specified, checking
for DTD conformity does not suffice for proof-checking.

\GAPT{}'s capabilities with respect to the {\ceresxml} format are asymmetric: \GAPT{} can read sequent calculus proofs
containing explicit permutation rules represented in the format, while it outputs (for technical reasons) sequent calculus
proofs containing implicit permutations. This asymmetry is due to backwards-compatibility with the \CERES{} system
and will be removed in the near future.


\subsection{Simple XML}

\simplexml is a simplified version of \ceresxml (i.e. its DTD has a much simpler structure and less restrictions, see Figure~\ref{fig:simple_dtd}) and is used to
represent arbitrary tree-like proofs (e.g. natural deduction, tableaux, etc). While \ceresxml fixes a certain language for
the formulas occuring in the proofs it encodes, the language of formulas is left open in \simplexml.
This allows the implementation of the proof parser (which follows the structure of the DTD) to be separated naturally from the
implementation of the formula parser (which is unrestricted in principle). This makes it very easy to exchange proofs
if proof checking is not desired, since then a generic prooftree-parser can be combined with a parser for the
language that is used in the proof. The {\simplexml} format was exploited to get a straightforward and easy implementation
of the parsing of proofs produced by {\RegSTAB}~\cite{regstab-ijcar2010}, which is a STAB prover, producing tableau refutations for formula schemata.

\begin{figure}[h]
\centering
\begin{tabular}{|c|}
\hline \\
\begin{minipage}{3.3in}
\begin{verbatim}
<!ELEMENT prooftrees	(proof*)>
<!ELEMENT proof		(rule)>
<!ATTLIST proof
          symbol	CDATA	#REQUIRED
          calculus	CDATA	#IMPLIED	>
<!ELEMENT rule
          (conclusion,(rule|prooflink)*)>
<!ATTLIST rule
          symbol	CDATA	#IMPLIED
          type		CDATA	#REQUIRED
          param		CDATA	#IMPLIED	>
<!ELEMENT conclusion	(#PCDATA)>
<!ELEMENT prooflink	EMPTY>
<!ATTLIST prooflink
          symbol	CDATA	#REQUIRED	>
\end{verbatim}
\end{minipage} \\ \\
\hline
\end{tabular}
 \caption{The full DTD file of \simplexml} \label{fig:simple_dtd}
 \end{figure}


There are disadvantages of using the {\simplexml} format: since it is a very general format, structural details about the nodes of the proof are absent. Therefore they are displayed simply as the strings which occur in the XML file instead of a more easily readable {\LaTeX} rendering. This also means some advanced features of {\pt} might not work. Basic features zooming and scrolling are not affected but certain views might not be available.

\subsection{Handy LKS}\label{sec:hlks}

Since proofs are the main input for the \GAPT{} system, a comfortable proof input language is important.
In the \CERES{} system, the Handy LK language\footnote{\url{http://www.logic.at/hlk/}} was used as that language,
and it allowed the input of proof schemata (by this we mean infinite sequences of proofs specified in an inductive way)
in a second-order language.

Unfortunately the Handy LK language and its implementation had several shortcomings.
First and foremost, it did not support full higher-order logic, and its implementation of proof schemata
was not based on a formal theory. In particular, there was no logical notion of a ``proof schema'' object --- only
the instances of the general schema were treated in a logical way.
Furthermore, the Handy LK compiler, which generated
a proof in \ceresxml from a proof written in Handy LK, was implemented separately from other parts of the
\CERES{} system, causing several practical problems.
Finally, the Handy LK compiler did not allow for a very fine-grained control over the resulting sequent calculus proofs, but
such control, while unimportant for the general task of formalizing proofs, can be important when experimenting with
proofs in the sense of structural proof-theory.


The replacement of Handy LK in GAPT is the \hlks language. This language is an implementation of the schematic proofs
defined and investigated in~\cite{Dunchev2012}.
The format is simple enough that an input proof can be written in any text editor. To differentiate it from other text files, the extension {\em .lks} is used. The full description of the grammar of the language as well as of the formal calculus, can be found in~\cite{Schemata}.

\begin{figure}[h]
\centering
\begin{tabular}{|l|}
 \hline \\
 \verb? proof \psi proves A(0), BigAnd(i=0..k) (~A(i) \/ A(i+1)) |- A(k+1) ? \\
 \verb? base { ? \\
 $\quad$ \verb? 1: autoprop(A(0), (~A(0) \/ A(1)) |- A(1)) ? \\
 $\quad$ \verb? root: andEqL3(1, (~A(0) \/ A(1)),  BigAnd(i=0..0) (~A(i) \/ A(i+1))) ? \\
 \verb? } ? \\
 \verb? step { ? \\
 $\quad$ \verb? 1: pLink((\psi, k) A(0), BigAnd(i=0..k) (~A(i) \/ A(i+1)) |- A(k+1)) ? \\
 $\quad$ \verb? 2: ax(A(k+1) |- A(k+1)) ? \\
 $\quad$ \verb? 3: negL(2, A(k+1)) ? \\
 $\quad$ \verb? 4: ax(A(k+2) |- A(k+2)) ? \\
 $\quad$ \verb? 5: orL(3, 4, ~A(k+1), A(k+2)) ? \\
 $\quad$ \verb? 6: cut(1, 5, A(k+1)) ? \\
 $\quad$ \verb? 7: andL(6, BigAnd(i=0..k) (~A(i) \/ A(i+1)), (~A(k+1) \/ A(k+2))) ? \\
 $\quad$ \verb? root: andEqL1(7, (BigAnd(i=0..k) (~ A(i) \/ A(i+1)) /\ ? \\
 $\hskip5em$ \verb? (~ A(k+1) \/ A(k+2))), BigAnd(i=0..k+1) (~ A(i) \/ A(i+1))) ? \\
 \verb? } ? \\ [1ex]
 \hline
\end{tabular}
  \caption{An example of a proof schema in \hlks.}
  \label{fig:hlks_proof}
\end{figure}

An {\em .lks} file must contain at least one proof definition, an example of which is given in Figure~\ref{fig:hlks_proof}. For an inductive proof definition, the \texttt{base} block describes the base case and the \texttt{step} block describes the recursive case. The IDs are arbitrary labels that are unique within the scope of \{ . . . \} blocks (i.e. the same labels can be used in the definition of base and step cases) and rules are tuples consisting of the rule's name, the IDs of the premises and of the auxiliary formulas. The \texttt{autoprop} keyword is used to prove propositional sequents automatically if the user is not interested in and does not want to give the exact proof of that sequent.

At the moment, {\hlks} is geared towards writing inductively defined sequences of proofs.
Of course,
non-inductively defined proofs (e.g.~just a single LK proof) can be trivially represented as inductively-defined proofs.
We plan to add syntactic sugar to the {\hlks} format to allow encoding of such simpler proofs in a natural way.

As the language would profit from syntax highlighting, having an editor for this language would be convenient. One solution is to use XText\footnote{XText homepage: \url{http://www.eclipse.org/Xtext/}} and create such an editor using the grammar of the \hlks language. An advantage of the grammar is that it is easy to give the exact line numbers where the parsing of a file fails because of a syntactic error.

\section{Features for Proof Mining}\label{sec:ProofMining} 

At the moment, \GAPT{} distinguishes two kinds of sequent-like proofs: first- and higher-order sequent calculus (LK) proofs and schematic first-order sequent calculus (LKS) proofs. In {\pt} there are separate menus for LK- and LKS-proofs, containing the possible operations for these proofs respectively. The available functionality from these menus are the following:

\begin{itemize}
 \item For LK-proofs:
   \begin{itemize}
    \item Compute the data-structures of the \CERES{} method, such as struct and characteristic clause set.
    \item Apply reductive cut-elimination (Gentzen's method). This option is available as well in the context menus of sequents occurring in the proof and applies cut-elimination to the proof ending in that sequent.
    \item Extract Herbrand sequent from a cut-free proof.
    \item Skolemize and/or regularize a proof.
   \end{itemize}

 \item For the LKS-proofs:
   \begin{itemize}
    \item Compute the data-structures of the schematic \CERES{} method, such as schematic struct, schematic characteristic clause set (see Figure~\ref{fig:pt_list}) and schematic projection term (see Figure~\ref{fig:pt_tree}).
    \item Compute the instance of a schematic proof, of a schematic struct or of a projection term for a particular number given by the user.
   \end{itemize}
\end{itemize}

Sometimes it is useful to get a list of the lemmas which are used in the proof. Therefore, there is a menu item extracting cut-formulas from LK and dedicated to LKS proofs.

The menus currently support only a subset of the capabilities of the \GAPT{} system. The other functionalities will be added to {\pt} on a by-need basis.

\section{Visualization Features}\label{sec:Visualization}

\begin{figure}[h]
\centering
  \includegraphics[scale=0.36]{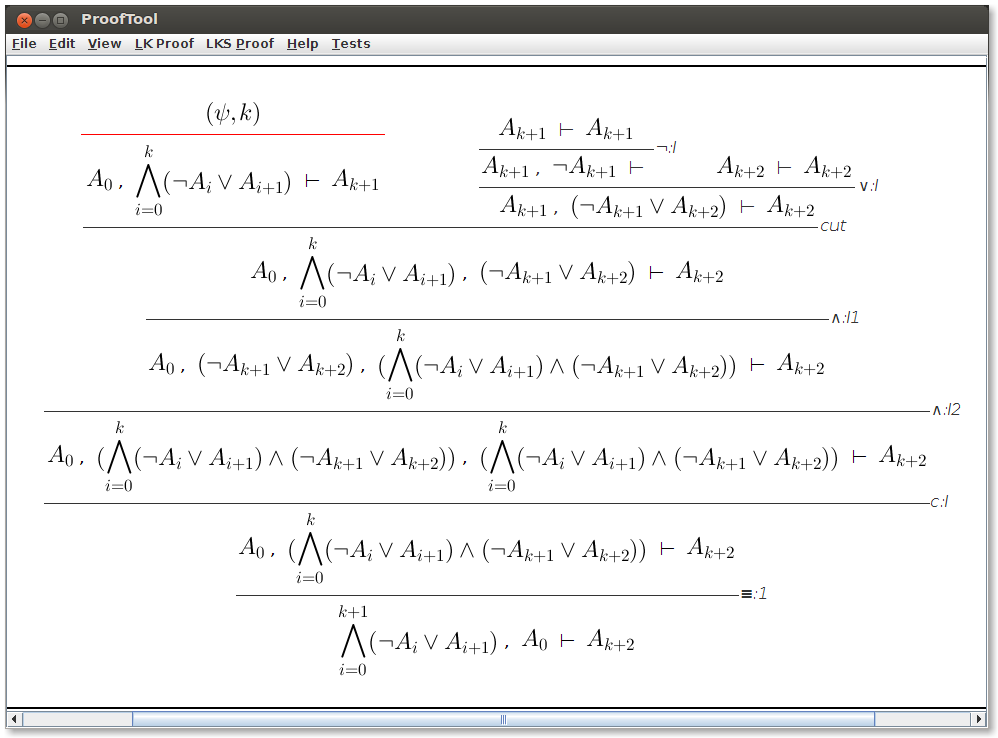}
  \caption{Proof in {\pt}.}
  \label{fig:pt_proof}
\end{figure}

{\pt} is a graphical user interface, used to display objects generated by the \GAPT{} system. These objects are: trees, proofs, sequents, formulas and the like. For example, the proof given in Figure~\ref{fig:hlks_proof} is displayed in {\pt} as shown in Figure~\ref{fig:pt_proof}.

A \textsc{TreeProof} is a binary tree in the \GAPT{} system which represents a tree-like proof. It is a \textsc{Tree} which is also a \textsc{Proof}. This means that we can expect that nodes are labeled with sequents. In general, proofs in the \GAPT{} system are sequent-like proofs. The reasons for the decision to display \textsc{TreeProof}s instead of \textsc{Proof}s are as follows: A \textsc{Proof} is only a directed acyclic graph, so additional arrows for shared structures would be necessary. Apart from sequent calculus proofs, at the moment only resolution proofs need to be displayed; The tree representations of these proofs can easily be obtained from the DAG-form.

The following features of {\pt} provide a better visualization of proofs:
\begin{description}
\item[Hide structural rules] has the purpose to shorten the size of proofs. In most of the cases structural rules do not contain any valuable information and they can be hidden for the user. The Edit$>$Hide Structural Rules menu item allows access to this feature.

\item[Hide sequent context] is used when the sequents occurring in a proof are very large. In such cases, for each inference in the proof, only its active formulas are shown.

\item[Mark cut-ancestors] highlights all ancestors of all cut-formulas occurring in the proof. It allows the user to have a glimpse of how cut-formulas are structured in the derivation.

\item[Split proof] allows focusing on a subproof of the proof. To shorten the size of the proof, it is also possible to hide/unhide some subproofs of it. These manipulations can be done from the context menus of the sequents.
\end{description}

\begin{figure}[h]
\centering
  \includegraphics[scale=0.36]{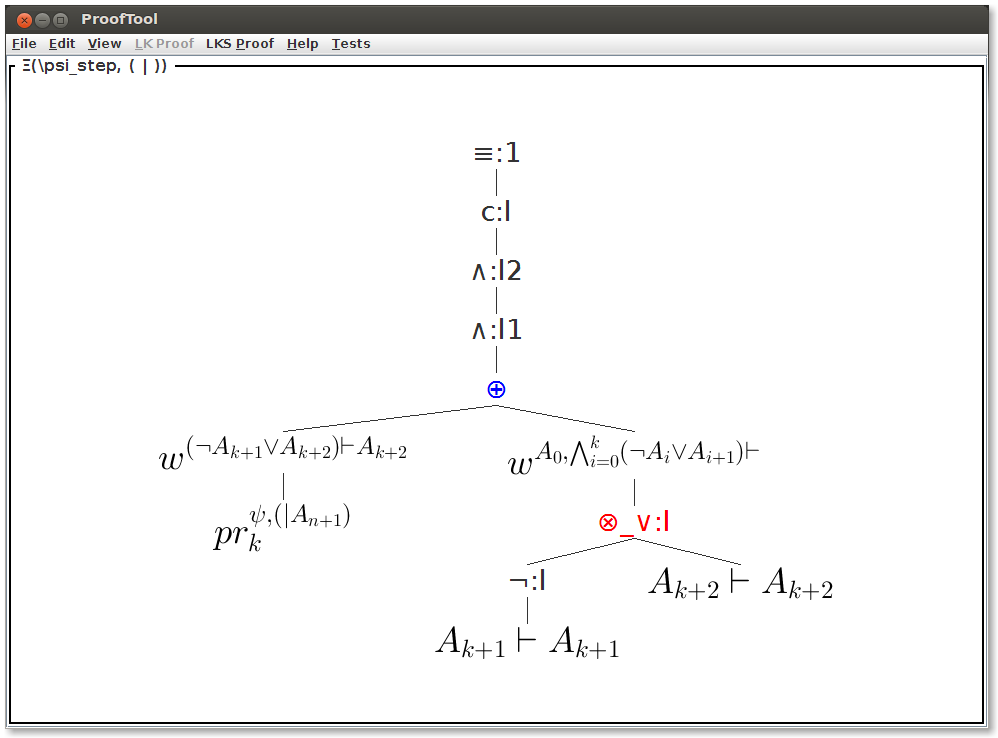}
  \caption{Tree in {\pt}.}
  \label{fig:pt_tree}
\end{figure}

\begin{figure}[ht]
\centering
\begin{minipage}[t]{0.40\linewidth}
\centering
\includegraphics[scale=0.4]{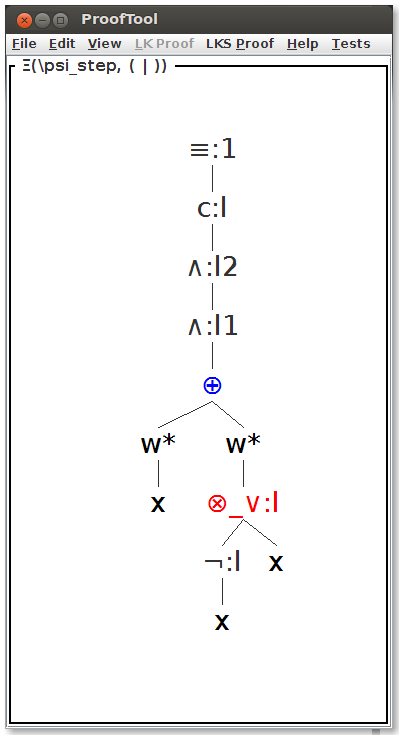}
  \caption{The same tree in {\pt} (leaves are hidden).}
  \label{fig:pt_tree_leaf}
\end{minipage}%
\hspace{0.08\linewidth}
\begin{minipage}[t]{0.40\linewidth}
\centering
 \includegraphics[scale=0.4]{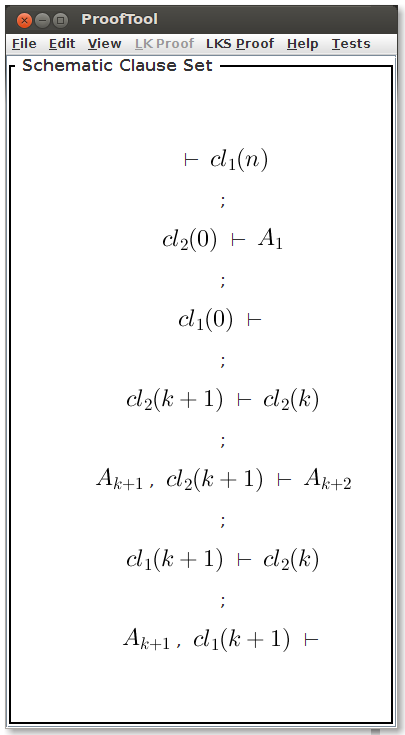}
  \caption{Sequent list in {\pt}.}
  \label{fig:pt_list}
\end{minipage}
\end{figure}

Unlike proofs, \textsc{Tree}s are binary trees in the system and they are displayed upside-down (see Figure~\ref{fig:pt_tree}). The user can manipulate the size of the tree by hiding/showing some branches or leaves. This is done by calling the corresponding menu items of the Edit menu, or by clicking on the vertex that should be hidden/shown (see Figure~\ref{fig:pt_tree_leaf}).

For tree-like objects like clause terms or projection terms, the system contains transformations to trees. A simple example is a directed acyclic graph where the corresponding tree just has duplicates of the shared structures.

Lists are very important data-structures and it is worthy to have specialized handling for them. In {\pt} each element of a list is displayed in a single line. Lines are separated with semicolons. The most commonly occurring lists in the \GAPT{} system are {\em sequent lists} and {\em definition lists}. For an example of a sequent list displayed using {\pt} see Figure~\ref{fig:pt_list}.

\begin{wrapfigure}{r}{0.4\textwidth}
\centering
  \includegraphics[scale=0.4]{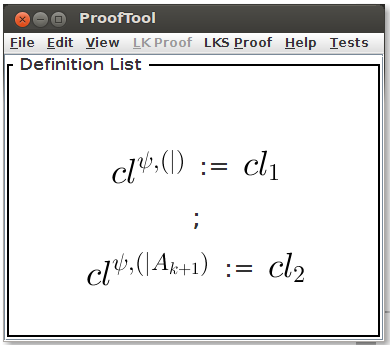}
  \caption{Definition list in {\pt}.}
  \label{fig:pt_list_defs}
\end{wrapfigure}

For the output to the user to be more readable, sometimes we abbreviate long expressions with shorter ones. These abbreviations are stored in the \GAPT{} system as a list of definitions. An example of the definition list for the clause set from Figure~\ref{fig:pt_list} is shown in Figure~\ref{fig:pt_list_defs}.

Thanks to a general design, which will be discussed in the next section, the basic features -- zooming and scrolling, can be applied to all objects easily. One more feature that is applicable to all the objects discussed above is searching.

\begin{description}
 \item[Search] is very useful when one has to deal with huge proofs or other objects, which is often the case in our research. In {\pt}, it is currently done in the following way: a user calls the Search dialog from the Edit menu and types the desired term (in {\LaTeX}). The desired term is then searched in the string representation of the displayed object and any occurrences found are highlighted in the displayed object. For example, when searching for \verb|cut| in the proof displayed in Figure~\ref{fig:pt_proof}, {\pt} finds the inference name and colors it green.

\end{description}

 A problem occurring during search is that the string representation of a node might be quite different from its rendering. When searching for the formula $\neg P_{k+1}$, the exact {\LaTeX} representation \verb*|\neg P_{k+1}| needs to be entered. Counter-intuitively, the search for \verb*|\neg  P_{k+1}| fails even if the rendered formula looks the same. To solve this problem, {\pt} allows the user to see the correct {\LaTeX} representation \verb*|\neg P_{k+1}| of (one of) the formulas $\neg P_{k+1}$ by right double clicking on it. The {\LaTeX} string can then be directly copied to the search dialog and all occurrences of $\neg P_{k+1}$ are highlighted. A drawback of this solution is that the user has to manually find an occurrence of the object he looks for. Avoiding this drawback would require approximate string matching~\cite{Bruno2007}, but for our current needs, this is not necessary.

\section{Implementation Details}\label{sec:id}

The \GAPT{} system is implemented in the programming language Scala. We rely heavily on functional features of Scala such as pattern matching. The parts of Scala's library which are of most used are the XML library, the combinator parser library and the Scala frontend for Java's SWING library. \textsc{jLatexMath}\footnote{\textsc{jLatexMath} homepage: \url{http://forge.scilab.org/index.php/p/jlatexmath/}} and \textsc{iText}\footnote{\textsc{iText} homepage: \url{http://www.itextpdf.com}} are the external libraries {\pt} depends on.

The domain specific language for XML that Scala provides reduces the parsing of our XML based input formats to pattern matching of XML tags and taking care of the differences in proof representations between \ceresxml{} and \GAPT{}'s internal datastructures. The remaining formats are read by combinator parsers which are also seamlessly integrated into the Scala language. Since most parts of our grammars are LL(1), we can benefit from annotations of non-backtracking rules but don't get much improvement of memoizing techniques as implemented in Scala's packrat parsers~\cite{packrat}.

Usage of the combinator parsers is shared with the \GAPT{}'s interactive theorem prover {\TAP} and the libraries connecting external theorem provers like {\Vampire}~\cite{Riazanov2002} and {\ProverN}~\cite{Prover9}. In the case of \ProverN{} the proof is first converted to the \IVY{} proof checker format by \ProverN{}'s \prooftrans{} utility. \IVY{} proofs are represented as Lisp S-expressions for which \GAPT{} utilizes combinator parsers for reading arbitrary S-expressions. Afterwards the resolution refutation is created from the \IVY{} specific structure by pattern matching S-expressions.

The PDF exporter in {\pt} is written using the \textsc{iText} library. It is a Java library for creating and manipulating the PDF files. We use its implementation of the abstract \textsc{Graphics2D} class to generate a {\em .pdf} file of the object drawn on the screen.

\begin{figure}[h]
\centering
  \includegraphics[scale=0.4]{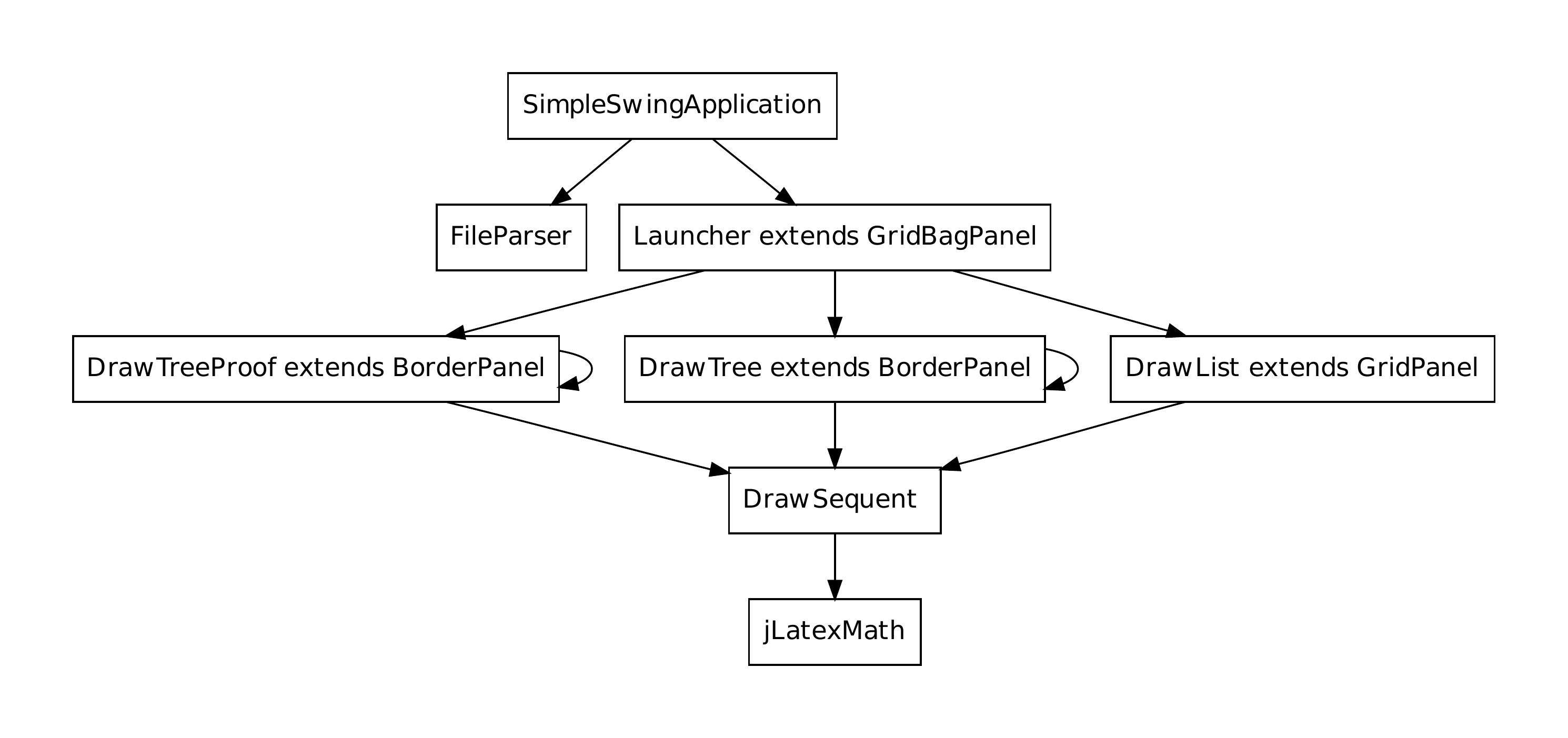}
  \caption{Architecture of {\pt}.}
  \label{fig:pt_architecture}
\end{figure}

{\pt} is implemented using the \textsc{scala.swing} library~\cite{swing_library}. It is a \textsc{SimpleSwingApplication} and consists of one frame, which contains a \textsc{MenuBar} and a \textsc{ScrollPane}. A brief description of the architecture of {\pt} is shown in Figure~\ref{fig:pt_architecture}.

\textsc{ScrollPane} has a component called \textsc{Launcher} which extends \textsc{GridBagPanel} and takes the following parameters:

\begin{itemize}
 \item A pair (\textsc{String}, \textsc{AnyRef}), where \textsc{AnyRef} is an object that should be displayed and \textsc{String} is a name of the object. The object has \textsc{TitledBorder} around it and the name is the title.
 \item An \textsc{Int}, which is a size of font that is used to display an object.
\end{itemize}

When an object is passed to \textsc{Launcher} it uses Scala's matching mechanism to recognize its structure and instantiates the corresponding class responsible for the drawing of the object. Basically, \textsc{Launcher} differentiates between three kinds of objects: trees, proofs and lists. The rendering classes are \textsc{DrawTree, DrawProof} and \textsc{DrawList} respectively.

\textsc{DrawTree} extends the \textsc{BorderPanel} class and displays a tree in the following way: A leaf node just renders the vertex, whereas an inner node draws the root node at the top and then creates new \textsc{DrawTree} instances for the child trees. For a binary node, the branches of the tree are rendered side-by-side on a frame and, for a unary node, the branch is rendered in the center. The root is connected to its children by straight lines.

\textsc{DrawProof} also extends the \textsc{BorderPanel} class and behaves similar to \textsc{DrawTree}. The difference is that the desired output looks like a sequent calculus proof. This means that while \textsc{DrawTree} puts the vertex at the top and the branches below, \textsc{DrawProof} puts the vertex at the bottom and the branches on top of it. Then it draws horizontal lines between vertices and puts the rule names next to the lines.

\textsc{DrawList} extends the \textsc{GridPanel} class having only one column and puts each element of the list in a separate cell.

In {\pt} a sequent is displayed by a \textsc{FlowPanel} which contains the representation of each formula as a separate \textsc{Label}.

To handle formulas it was decided to use \textsc{jLatexMath}. It is a Java API which displays mathematical formulas written in {\LaTeX} as images, which was also used in the GUI of Scilab\footnote{Scilab homepage: \url{http://www.scilab.org/}}. The drawback of this library is that having an image for every formula is quite memory consuming.

For each formula displayed a label is created. Then the formula is transformed into a {\LaTeX} string and rendered using \textsc{jLatexMath}. The resulting image is assigned as an icon to the label. Since this rendering is expensive, we use it only when necessary and display simple string representations otherwise. For example,  any vertex containing a higher-order expression of type \textsc{HOLExpression} (which also includes first-order and schema formulas) is rendered by \textsc{jLatexMath}. Other types of vertices are displayed simply as their Scala string representation which often suffices since the Unicode character-set used includes Greek letters as well as other logical symbols.

{\pt} strongly profits from the \textsc{scala.swing} wrapper library which simplifies event handling significantly. Since there are only \textsc{Publisher} and \textsc{Reactor} elements instead of Java's complicated event handling mechanism, for instance menu items only need to listen to the \textsc{ProofToolPublisher} to adjust their activation status accordingly. In the case a new file is loaded or the proof database is dynamically changed the  \textsc{ProofDbChanged} event is issued. Since also the View$>$View Proof, View$>$View Clause List and View$>$View Term Tree menus listen to this event, they can refresh their list of \textsc{MenuItem}s.

\section{Conclusions and Future Work}\label{sec:conclusion}

In the previous sections, several features of {\pt} were described. These features are summarized in Figure~\ref{fig:pt_vs_others}. {\pt}'s predecessor, known as {\oldpt}, is also shown. In comparison, it is noticeable that {\pt}'s feature set is significantly broader than {\oldpt}'s. While {\pt} can display general tree-like structures, such as structs and projection terms, {\oldpt} can only display sequent calculus proofs. Even when {\oldpt} displays a list of sequents, for instance, what is displayed is actually a fake proof with dummy unary inferences connecting the sequents. Moreover, {\pt} accepts more input and output proof formats than {\oldpt}. Most importantly, proof schemata can now be handled and displayed directly, while {\oldpt} could only display traditional sequent calculus proofs, and hence only particular instances of proof schemata. Finally, while {\oldpt} was just a proof viewer, {\pt} gives access to \GAPT{}'s proof mining algorithms, such as cut-elimination and Herbrand sequent extraction.
\begin{figure}[h]
\centering
\begin{tabular}{| c | l@{} | c | c | c | c |}
 \hline
\multicolumn{2}{|c|}{\textbf{Features}}& \pt & \oldpt & \LOUI & \IDV \\
\hline
\multirow{6}{*}{Input/Output}& Parse ceres XML & \Yes & \Yes & \No & \No \\
\cline{2-6}
& Parse simple XML & \Yes & \No & \No & \No \\
\cline{2-6}
& Parse TPTP/TSTP & * & \No & \Yes & \Yes \\
\cline{2-6}
& Export to TSTP & * & \No & \No & \No \\
\cline{2-6}
& Export to .tex & \Yes & \Yes & \No & \No \\
\cline{2-6}
& Export to .tptp & \Yes & \No & \Yes & \Yes \\
\cline{2-6}
& Export to .pdf & \Yes & \No & \No & \No \\
\hline
\multirow{7}{*}{Visualization}& Zooming, scrolling & \Yes & \Yes & \Yes & \Yes \\
\cline{2-6}
& Sequent calculus proofs & \Yes & \Yes & \No & \No \\
\cline{2-6}
& Sequent lists & \Yes & \Yes & \No & \No \\
\cline{2-6}
& Definition lists & \Yes & \No & \No & \No \\
\cline{2-6}
& Proof schemata & \Yes & \No & \No & \No \\
\cline{2-6}
& Trees and related features & \Yes & \No & \Yes & \No \\
\cline{2-6}
& DAGs and related features & * & \No & \Yes & \Yes \\
\hline
\multirow{10}{*}{Proof Mining}& Marking (cut-)formula ancestors & \Yes & \No & \No & \Yes \\
\cline{2-6}
& Extracting cut-formulas & \Yes & \No & \No & \No \\
\cline{2-6}
& Extracting Herbrand sequent & \Yes & \No & \No & \No \\
\cline{2-6}
& Hiding sequent context & \Yes & \No & \No & \No \\
\cline{2-6}
& Hiding structural rules & \Yes & \No & \No & \No \\
\cline{2-6}
& Search & \Yes & \No & \No & \No \\
\cline{2-6}
& Split/unsplit & \Yes & \Yes & \Yes & \Yes \\
\cline{2-6}
& Substitute/unsubstitute & \No & \Yes & \No & \No \\
\cline{2-6}
& Proofs in natural language & \No & \No & \Yes & \No \\
\cline{2-6}
& Cut-elimination (Gentzen, \CERES{}) & \Yes & \No & \No & \No \\
\cline{2-6}
& Skolemization, regularization & \Yes & \No & \No & \No \\
\hline
\end{tabular}
  \caption{Features implemented by {\pt} and other similar systems.}
  \label{fig:pt_vs_others}
\end{figure}

{\LOUI} and {\IDV} are also shown, but care should be taken when using this table for comparison. As the table is focused on features available in {\pt} and relevant for \GAPT{}'s needs, it is most probably the case that this table is lacking features implemented by {\LOUI} or {\IDV}, but not by {\pt}. Moreover, the fact that {\IDV} and {\LOUI} do not implement some listed features implemented by {\pt} should not be considered as a flaw of {\IDV} and {\LOUI}, since most of these features are very specific to \GAPT{}'s needs. Nevertheless, the table does include a few of {\IDV}'s or {\LOUI}'s features that are relevant for \GAPT{} but not yet implemented by {\pt}. Those marked with ``*'' will be the focus of imminent future work.

%

Besides the implementation of features marked with ``*'' in the table, another task that remains for the future is the improvement of the {\hlks} language aiming at unifying parsing of schematic and traditional sequent calculus proofs. Furthermore, there is plenty of room for decreasing the amount of detail that {\hlks} requires in the proof specifications, beyond what is currently possible with the ``autoprop'' feature. In particular, the next version of {\hlks} intends to additionally allow the omission of structural inferences such as contraction and weakening, which are poor in mathematical content.


\section*{Acknowledgements}

We would like to thank the anonymous referees and all the participants of the 10th Workshop on User Interfaces for Theorem Provers for their useful comments and suggestions about this work. They helped us to improve not only the contents and presentation of this paper, but also the {\pt} program.

\bibliographystyle{eptcs}
\bibliography{bibliography}

\end{document}